\documentclass[prl,aps,floatfix,showpacs,twocolumn]{revtex4}
\usepackage{dcolumn}
\usepackage{bm}
\usepackage{color}
\usepackage{graphicx}
\usepackage{epsf}
\usepackage{pstricks}
\begin{document}
\title{The proton-proton weak capture in chiral effective field theory}
\author{L.E.\ Marcucci$^{\, {\rm a,b}}$, 
R.\ Schiavilla$^{\,{\rm c,d}}$, and M.\ Viviani$^{\,{\rm b}}$}
\affiliation{
$^{\,{\rm a}}$\mbox{Department of Physics, University of Pisa, 56127 Pisa, Italy}\\
$^{\,{\rm b}}$\mbox{INFN-Pisa, 56127 Pisa, Italy}\\
$^{\,{\rm c}}$\mbox{Department of Physics, Old Dominion University, Norfolk, VA 23529, USA}\\
$^{\,{\rm d}}$\mbox{Jefferson Lab, Newport News, VA 23606}\\
}

\date{\today}

\begin{abstract}
The astrophysical $S$-factor for proton-proton weak
capture is calculated  in chiral effective field theory over the
center-of-mass relative-energy range 0--100 keV.  The chiral
two-nucleon potential derived up to next-to-next-to-next-to
leading order is augmented by the full electromagnetic interaction
including, beyond Coulomb, two-photon and vacuum-polarization
corrections.  The low-energy constants (LEC's) entering the weak
current operators are fixed so as to reproduce the $A=3$ binding
energies and magnetic moments, and the Gamow-Teller matrix
element in tritium $\beta$ decay.  Contributions from $S$ and $P$
partial waves in the incoming two-proton channel are retained.
The $S$-factor at zero energy is found to be $S(0)=(4.030
\pm 0.006)\times 10^{-23}$ MeV fm$^2$,
with a $P$-wave contribution of $0.020\times 10^{-23}$
MeV fm$^2$.  The theoretical uncertainty is due to the fitting
procedure of the LEC's and to the cutoff dependence.
\end{abstract}


\index{}\maketitle
The proton weak capture on protons, {\it i.e.}, the reaction
$^1$H$(p,e^+\nu_e)^2$H (hereafter labelled $pp$), is the most
fundamental process in stellar nucleosynthesis: it is the first
reaction in the $pp$ chain, which converts hydrogen into helium
in main sequence stars like the Sun.  Its reaction rate is
expressed in terms of the astrophysical $S$-factor, $S(E)$,
where $E$ is the two-proton center-of-mass (c.m.) energy.  At the
center of light stars like the Sun, with temperature of the order of
$1.5\times 10^7$ K, the Gamow peak is at $E \simeq 6$ keV.
At these energies, the reaction cross section cannot be measured
in terrestrial laboratories, and it is necessary to rely on theoretical
predictions, which are typically given for $S(0)$---the zero-energy
value of the $S$-factor. The many studies on $S(0)$
have been extensively reviewed in Ref.~\cite{SFII}, and are
succinctly summarized next.

The currently recommended value for $S(0)$, $(4.01 \pm 0.01)
 \times 10^{-23}$ MeV fm$^2$~\cite{SFII}, is the average
 of values obtained within three different approaches, labelled
``potential models'' (PM), ``hybrid chiral effective field theory''
($\chi$EFT*) and ``pionless effective field theory''
($\setminus$\hspace*{-0.2cm}$\pi$EFT).  The first one uses
phenomenological realistic models for the nuclear potential,
fitted to reproduce the large body of two-nucleon ($NN$) bound
and scattering state data with a $\chi^2/$datum $\sim 1$. 
The axial current operator includes both one-body terms,
determined from the coupling of the single nucleon to the weak
probe, and two-body terms, derived from meson-exchange
mechanisms and the excitation of $\Delta$-isobar resonances.
These two-body terms are constrained to reproduce the
experimental value of the Gamow-Teller (GT) matrix element of
tritium $\beta$-decay.

In the hybrid approach, transition operators derived in
$\chi$EFT are sandwiched between initial and final wave
functions generated by potential models. The only unknown
low-energy constant (LEC), which parametrizes the strength
of a contact-type four-nucleon coupling to the axial current,
is determined by fitting the experimental GT matrix element.  

Finally, $\setminus$\hspace*{-0.2cm}$\pi$EFT is an effective 
field theory approach applicable to low-energy processes---such
as the $pp$ reaction under consideration here---with a
characteristic momentum $Q$ much smaller than the pion mass
$m_\pi$.  In such a theory, pions are integrated out and the $NN$
interaction and weak currents are described by classes of
point-like contact interactions, each class corresponding to given
order in a systematic expansion in powers of $Q/m_\pi$.

The energy-dependence of $S(E)$ in the $pp$ capture (and
other reactions as well in the $pp$ chain) is often parametrized
as~\cite{SFII}
\begin{equation}
S(E)=S(0)+S^\prime(0)E+S^{\prime\prime}(0)E^2/2+\cdots \ ,
\label{eq:taylor}
\end{equation}
where $S^\prime(0)$ and $S^{\prime\prime}(0)$ are the first and 
second derivatives of $S(E)$, evaluated at $E=0$.  For
$S^\prime(0)$ and  $S^{\prime\prime}(0)$ the situation is
much less clear than for $S(0)$. The adopted value for
$S^\prime(0)$ in Ref.~\cite{SFII} is $S^\prime(0)/S(0)=(11.2
\pm 0.1)$ MeV$^{-1}$, as obtained in Ref.~\cite{Bac69} and
later confirmed in Ref.~\cite{Sch98} in a PM approach.
No value is reported for $S^{\prime\prime}(0)$ in Ref.~\cite{SFII}.
In Ref.~\cite{Bac69} it was estimated by dimensional
considerations that the contribution of $S^{\prime\prime}(0)$
to the $pp$ rate would be at the level of 1\% for temperatures
characteristic of the solar interior.  Only very recently, $S^\prime(0)$
and $S^{\prime\prime}(0)$ have been calculated in
$\setminus$\hspace*{-0.2cm}$\pi$EFT~\cite{Che12}
to the third-order in the power expansion with the results
$S^\prime(0)/S(0)=(11.3\pm 0.1)$ MeV$^{-1}$ and
$S^{\prime\prime}(0)/S(0)=(170\pm 2)$ MeV$^{-2}$.
In conclusion, a systematic study of $S(E)$ in 
(pionfull) $\chi$EFT is still missing.  We address this
omission in the present letter.

The $NN$ potential is that derived in $\chi$EFT up to
next-to-next-to-next-to leading order (N3LO) in the chiral
expansion by Entem and Machleidt~\cite{Ent03,Mac11}.
However, in the $pp$ sector, it is augmented by
the inclusion of higher-order electromagnetic terms, due
to two-photon exchange and vacuum polarization.
These higher-order terms are the same as those of the
Argonne $v_{18}$ (AV18) $NN$ potential~\cite{Wir95}, and
therefore also retain short-range corrections associated with the
finite size of the proton charge distribution.  The additional
distortion of the $pp$ wave function, induced primarily by
vacuum polarization, has been shown to reduce $S(0)$ by
$\sim 1$\% in Ref.~\cite{Sch98}.

\begin{figure}[t]
\includegraphics[width=2.5in,height=0.8in]{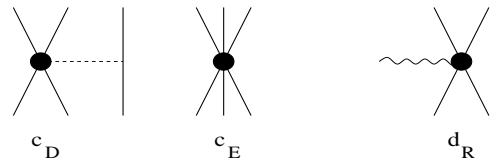}
\caption{
One-pion exchange plus $NN$ contact, and $NNN$ contact
terms entering the three-nucleon potential at N2LO, and the
contact term entering the $NN$ axial current.  Solid, dashed,
and wavy lines represent, respectively, the nucleon, pion, and
external probe.}
\label{fig:contact}
\end{figure}

The charge-changing weak current
has been derived up to N3LO in Ref.~\cite{Par96ecc}.  Its
polar-vector part is related, via the conserved-vector-current
constraint, to the (isovector) electromagnetic current, and
includes, apart from one- and two-pion-exchange terms,
two contact terms---one isoscalar and the other isovector---whose
strengths are parametrized by the LEC's $g_{4S}$ and $g_{4V}$.
The two-body axial-vector current includes terms of one-pion 
range as well as a single contact current, whose strength is
parametrized by the LEC $d_R$.  The latter is related to the LEC 
$c_D$, which, together with $c_E$, enters the three-nucleon
($NNN$) potential at next-to-next-to leading order (N2LO),
as illustrated in Fig.~\ref{fig:contact}.

These chiral potentials and currents have power-law behavior in
momentum space, and must be regularized before they can be
used in practical calculations.  This is accomplished by multiplying
them by a momentum-cutoff function, whose cutoff $\Lambda$ is
taken to be in the range (500--600) MeV.  Finally, we should note
that inclusion of such a cutoff spoils the requirement of
conserved-vector and partially-conserved-axial currents.
In particular, we note that the construction of a conserved vector
current with the N3LO $NN$ potential used here would
require accounting for two-loop corrections, a task
well beyond the present state of the art.

The LEC's $c_D$ (or $d_R$), $c_E$, $g_{4S}$ and $g_{4V}$
are determined with the procedure discussed in
Ref.~\cite{Mar12}.  First, the values of the LEC's $\{c_D,c_E\}$
which reproduce the $A=3$ binding energies are obtained for
both $\Lambda=500$ and 600 MeV, with $c_D$ in the range
$(-3,3)$.  Next, within this range, the GT matrix
element is calculated and $c_D$ (or equivalently $d_R$) is fixed
to reproduce its experimental value.  The range of $c_D$ values,
for which the calculated GT matrix element is within the lower and
upper limits of its experimental determination, are $(-0.20, -0.04)$
for $\Lambda=500$ MeV, and $(-0.32,-0.19)$ for $\Lambda=600$
MeV.  The corresponding ranges for $c_E$ are $(-0.208, -0.184)$
and  $(-0.857, -0.833)$, respectively~\cite{Mar12}.  Lastly, for the
minimum and maximum values of $\{c_D,c_E\}$ and the given
$\Lambda$, the isoscalar and isovector LEC's $g_{4S}$ and
$g_{4V}$ are determined by reproducing the $A=3$ magnetic
moments.  The values for all the LEC's are listed in Table I of
Ref.~\cite{Mar12}.  Indeed, in that work it was shown that the
consistent $\chi$EFT approach outlined above leads to predictions
(with an estimated theory uncertainty of about 1\%) for the rates of
muon capture on deuteron and $^3$He, that are in excellent
agreement with the experimental data.

All earlier studies of the $pp$ capture we are aware of (see
Ref.~\cite{SFII} and references therein) have only considered
the $^1S_0$ channel in the initial $pp$ scattering state.  Since one
of the objectives of the present work is to study the energy
dependence of the $S$-factor up to $E$=100 keV,
we include, in addition to the $^1S_0$, the $P$-wave channels
$^3P_0$, $^3P_1$, and $^3P_2$.  We outline the calculation
in the following, deferring a more extended discussion of it to a
later paper~\cite{Marinprep}.  
 
The $pp$ weak capture cross section $\sigma(E)$, from which
the $S$-factor is obtained as 
$S(E)=E\, {\rm exp}(2\pi\,\eta) \, \sigma(E)$ ($\eta=\alpha/v_{\rm rel}$,
$\alpha$ being the fine structure constant and
$v_{\rm rel}$ the $pp$ relative velocity), 
is written in the c.m. frame as
\begin{eqnarray}
&&\sigma(E)=\int 2\pi\delta(\Delta m + E -\frac{q^2}{2m_d} 
-E_e - E_\nu) \frac{1}{v_{\rm rel}}\nonumber \\
&\times&F(Z,E_e) \frac{1}{4}
\sum_{s_e\,s_\nu}\sum_{s_1\,s_2\,s_d} |\langle f|H_W|i \rangle |^2 
\frac{d{\bf p}_e}{(2\pi)^3}\frac{d{\bf p}_\nu}{(2\pi)^3}\ , 
\label{eq:xs}
\end{eqnarray}
where $\Delta m = 2\, m_p-m_d$ ($m_p$ and $m_d$ are the
proton and deuteron masses, respectively), ${\bf p}_e$
(${\bf p}_\nu$) and $E_e$ ($E_\nu$) are the electron (neutrino)
momentum and energy, ${\bf q}={\bf p}_e+{\bf p}_\nu$ is the 
momentum transfer, and
$F(Z,E_e)$ is the Fermi function (with $Z=1$), which accounts for
the Coulomb distortion of the final positron wave function.  Its
explicit expression can be found in Ref.~\cite{She12}, increased
by 1.62\% to take into account radiative corrections to the cross
section~\cite{Kur03}.  The transition amplitude is given by
\begin{equation}
\langle f |H_W|i \rangle = \frac{G_V}{\sqrt{2}}l^\sigma 
\langle -{\bf q}; d |j_\sigma^\dagger |{\bf p}; pp \rangle\ ,
\label{eq:trans}
\end{equation}
where $G_V$ is the Fermi constant ($G_V=1.14939\times10^{-5}$ 
GeV$^{-2}$~\cite{Har90}), $| -{\bf q}; d \rangle$ and
$|{\bf p}; pp \rangle$ represent, respectively, the deuteron
bound state with recoiling momentum $-{\bf q}$ and the $pp$ 
scattering state with relative momentum ${\bf p}$, and
$l_\sigma$ and $j^\sigma({\bf q})$ 
are the leptonic and nuclear weak currents, respectively.
A standard multipole decomposition of the nuclear
weak current operator leads to~\cite{Mar00}
\begin{equation}
\frac{1}{4}\sum_{s_e\,s_\nu}\sum_{s_1\,s_2\,s_d} |\langle f|H_W|i \rangle |^2
=
(2\pi)^2G_V^2L_{\sigma\tau}N^{\sigma\tau} \ ,
\label{eq:hhw}
\end{equation}
where the lepton tensor $L^{\sigma\tau}$ is written in terms of
electron and neutrino four velocities,
and the nuclear tensor is defined as
\begin{equation}
N^{\sigma\tau}= \sum_{s_1\,s_2\,s_d}
W^\sigma({\bf q}; s_1 s_2 s_d)\,W^{\tau *}({\bf q}; s_1 s_2 s_d)\ ,
\label{eq:nst}
\end{equation}
with
\begin{eqnarray}
W^{\sigma=0,3}({\bf q}; s_1 s_2 s_d)\!\!&=&\!\!\!\!\!\!
\sum_{LSJ; \Lambda\geq 0} \!\!\!\!\!\!X_0^{LSJ\Lambda}({\hat{\bf q}}; s_1 s_2 s_d) 
T_\Lambda^{LSJ}(q) , 
\label{eq:w03} \\
W^{\sigma=\lambda}({\bf q}; s_1 s_2 s_d)&=&
-\frac{1}{\sqrt{2}} 
\sum_{LSJ; \Lambda\geq 1} X_{-\lambda}^{LSJ\Lambda}({\hat{\bf q}}; s_1 s_2 s_d) 
\nonumber \\
&\times&[\lambda M_\Lambda^{LSJ}(q) + E_\Lambda^{LSJ}(q)] \ , 
\label{eq:wpm}
\end{eqnarray}
where $\lambda=\pm 1$ denote spherical components.
The spin-quantization axis for the hadronic states is taken along
the direction ${\hat {\bf p}}$ of the $pp$ relative momentum.  The
functions $X_{\lambda=0,\pm 1}^{LSJ\Lambda}({\hat{\bf q}};
s_1 s_2 s_d)$ depend on the direction ${\hat{\bf q}}$, the proton
and deuteron spin projections $s_1$, $s_2$ and $s_d$, and
we have used the notation
$T_\Lambda^{LSJ}(q)=C_\Lambda^{LSJ}(q)$ or $L_\Lambda^{LSJ}(q)$ 
for $\sigma=0$ or 3. The quantities $C_\Lambda^{LSJ}(q)$,
$L_\Lambda^{LSJ}(q)$, $M_\Lambda^{LSJ}(q)$ and
$E_\Lambda^{LSJ}(q)$ are, respectively, 
the reduced matrix elements (RME's) for the
Coulomb, longitudinal, transverse magnetic and transverse electric
multipole operators between the initial $pp$ state with
orbital angular momentum $L$, channel spin $S$ ($S=0,1$),
total angular momentum $J$, and the final deuteron state
with total angular momentum $J_d=1$.
The number $\Lambda$ in Eqs.~(\ref{eq:w03}) and~(\ref{eq:wpm})
is the multipole order, with 
${\bf{\Lambda}}+{\bf{J}}={\bf{J_d}}$.
The integrations over ${\bf p}_e$ and
${\bf p}_\nu$ are performed by Gaussian
quadratures~\cite{Mar00}, and a moderate
number of Gauss points ($\sim$ 10--20 for
each integration) suffices to achieve
convergence to within better than 1 part in $10^3$.

The two-body wave functions corresponding to
the non-local chiral potentials of Refs.~\cite{Ent03,Mac11}
have been obtained variationally with the technique
described in Ref.~\cite{Mar11}.  In the present work,
there is the complication due to the presence,
in the $pp$ sector, of higher-order corrections
(from two-photon exchange and vacuum polarization)
in the electromagnetic potential $v_{em}(r)$.
We proceed in the following way.  We first calculate the
regular and irregular solutions corresponding to $v_{em}(r)$ only
by direct integration of the the Schr\"odinger equation---these are
denoted as $\Omega^{(R)}$ and $\Omega^{(I)}$.
We then expand the $pp$ continuum wave function in
channel $\alpha\equiv LSJJ_z$ as
\begin{equation}
\Psi^\alpha=\sum_\mu c^\alpha_\mu \Psi^\alpha_\mu + 
\Omega_\alpha^{(R)}
+\sum_{\alpha^\prime}R_{\alpha \alpha^\prime}\,\Omega_{\alpha^\prime}^{(I)}\ ,
\label{eq:psi}
\end{equation}
where $\Psi^\alpha_\mu$ are known functions written 
as product of Laguerre polynomials
(see Eq.~(3.1) of Ref.~\cite{Mar11}), which vanish at large
inter-particle separations.  
Clearly, the dependence on the $NN$ potential enters only in
the unknown coefficients $c_{\mu}$
and matrix elements $R_{\alpha \alpha^\prime}$, which are
determined via the Kohn variational principle.
A system of linear inhomogeneous equations for the $c_{\mu}$'s and a set
of algebraic equations for the $R_{\alpha \alpha^\prime}$'s result,
which are solved by standard techniques.  From the
$R_{\alpha \alpha^\prime}$'s, phase shifts and mixing angles are
easily obtained.  We have verified that, in the case of the AV18,
the method outlined above leads to $^1S_0$ phase shifts
in agreement with those reported for the AV18 in Ref.~\cite{Wir95}
(which included the same $v_{em}(r)$ used here).
We have also verified that we are able to reproduce
the N3LO phase shifts of Ref.~\cite{Mac11},
obtained by including only the Coulomb potential 
in $v_{em}(r)$.
Further details will be reported in a later
publication~\cite{Marinprep}.

The cumulative $S$- and $P$-wave contributions
to the astrophysical $S$-factor at zero energy are listed in
Table~\ref{tab:s0}.  Inspection of the table shows that:
(i) the cutoff dependence is negligible as is the overall theoretical
uncertainty (well below 1\%) due to the procedure adopted to
fit the LEC's entering the current; (ii) the $P$-wave
contributions to $S(0)$ sum up to $\sim$ 1\% of the total 
value; (iii) the results can be summarized in the conservative
range $S(0)=(4.030 \pm 0.006)\times 10^{-23}$ MeV fm$^2$.
For comparison, we have also calculated $S(0)$ within the
PM approach, using the AV18 potential and the same model for
the nuclear current of Refs.~\cite{Mar00,Mar11,Sch98},
obtaining $S(0)=(4.033\pm 0.003)\times 10^{-23}$ MeV fm$^2$
($S(0)=(4.000\pm 0.003)\times 10^{-23}$ MeV fm$^2$) when
all the $S$- and $P$-waves (only the $^1S_0$ channel) are
included. The agreement between the PM and $\chi$EFT
results is excellent.  Finally, it should be noted that the $^1S_0$
$S$-factor, in units of $10^{-23}$ MeV fm$^2$, obtained with the
pure Coulomb interaction, is 4.025 when $\Lambda=500$ MeV, 
and 4.030 within the PM approach with the AV18.  Therefore,
while the full electromagnetic interaction accounts for a 
$\sim 1$\% reduction in $S(0)$, this effect is 
in practice offset by the $P$-wave contributions.
\vspace*{-0.5cm}
\begin{table}[tbh]
\caption{Cumulative $S$- and $P$-wave contributions
to the astrophysical $S$-factor at zero c.m.
energy in units of $10^{-23}$ MeV fm$^2$.  The theoretical
uncertainties are given in parentheses and are due to the fitting
procedure adopted for the LEC's in the weak current.  The results
have been obtained with the two different cutoff values 
$\Lambda=500$ and 600 MeV.}
\label{tab:s0}
\begin{tabular}{c|cccc}
 & $^1S_0$ & $^3P_0$ & $^3P_1$ & $^3P_2$ \\
\hline
$\Lambda$=500 MeV & 4.008(5) & 4.011(5) & 4.020(5) & 4.030(5) \\
$\Lambda$=600 MeV & 4.008(5) & 4.010(5) & 4.019(5) & 4.029(5) \\
\hline
\end{tabular}
\end{table}
\begin{figure}[tbh]
\includegraphics[width=2.5in,height=1.5in]{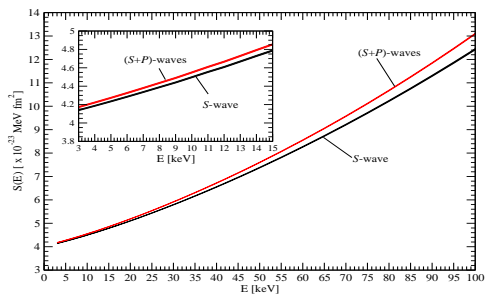}
\caption{(Color online) 
The astrophysical $S$-factor as function of the c.m.
energy in the range 0--100 keV.  The $S$- and
$(S+P)$-wave contributions are displayed separately.  In
the inset, $S(E)$ is shown in the range 3--15 keV.}
\label{fig:se}
\end{figure}

The energy dependence of $S(E)$ is shown in Fig.~\ref{fig:se}.
The $S$- and $(S+P)$-wave contributions are displayed separately, and
the theoretical uncertainty is included---the curves are in fact very
narrow bands.  As expected, the $P$-wave contributions become
significant at higher values of $E$. 

%
\begin{table}[tbh]
\caption{Values for 
$S^{n}(0)/S(0)$ with $n=1$--4, 
in units of MeV$^{-n}$, and 
the $\chi^2$ as defined in the text,
obtained with a polynomial fit of $S(E)$ up to orders $O(E^2)$ 
(Fit 1), $O(E^3)$ (Fit 2), and $O(E^4)$ (Fit 3),
retaining all $(S+P)$-waves.
The results obtained by retaining
only the $S$ channel are
listed separately for Fit 1 and 2. 
Also listed are the results of Ref.~\protect\cite{Che12}.
The theoretical uncertainties, listed only for $n=1,2,3$,
are given in parentheses and account for the cutoff sensitivity
and the errors due to the LEC's fitting procedure.}
\label{tab:fit}
\begin{tabular}{c|ccccc}
 $n$ & $1$ & $2$ & $3$ & $4$ & $\chi^2$\\
\hline
Fit 1 & 12.59(1) & 199.3(1) &             &                  & 8.8$\times 10^{-4}$ \\
Fit 2 & 11.94(1) & 248.8(2) & --1183(8)   &                  & 1.9$\times 10^{-4}$ \\
Fit 3 & 11.34(1) & 327.1(5) & --5592(12)  & 99 $\times 10^3$ & 2.0$\times 10^{-5}$ \\
\hline 
$S$ - Fit 1 & 12.23(1) & 178.4(3) &           & & 1.2$\times 10^{-3}$ \\
$S$ - Fit 2 & 11.42(1) & 239.6(5) & --1464(5) & & 1.9$\times 10^{-4}$ \\
\hline
$S$ - Ref.~\protect\cite{Che12} 
                & 11.3(1)  & 170(2)   &       & & 3.4$\times 10^{-1}$ \\
\hline
\end{tabular}
\end{table}

\begin{figure}[bth]
\includegraphics[width=2.5in,height=1.5in]{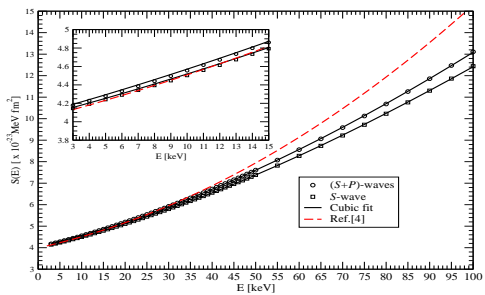}
\caption{(Color online) The astrophysical $S$-factor
in the energy range 0--100 keV, calculated with
$\Lambda=500$ MeV and $c_D=-0.20$ and including
the $S$- and $(S+P)$-wave contributions is
compared with the cubic polynomial fit and the quadratic fit
of Ref.~\protect\cite{Che12}.  In the inset, $S(E)$ is shown
in the range 3--15 keV.}
\label{fig:fit}
\vspace*{-0.3cm}
\end{figure}
Next, we examine the question of whether the polynomial
approximation for $S(E)$ given in Eq.~(\ref{eq:taylor}) is justified.
To this end, we have performed a least-squares polynomial fit to $S(E)$ up to
order $O(E^2)$, {\it i.e.}, by using Eq.~(\ref{eq:taylor}) itself, and
up to order $O(E^n)$, by adding terms
$S^{n}(0)E^n/n!$, with $n=3,4$ ($S^{n}(0)$ is the $n$-th
derivative of $S(E)$ evaluated at $E=0$).  The values
for $S^{n}(0)$, with $n=1$--4, are listed in Table~\ref{tab:fit},
along with the $\chi^2$ value, which we define as
the sum over all the energy grid values of the ``normalized''
residuals, $\chi^2=\sum_i(1-f_i^{fit}/f_i^{calc})^2$, where $f_i^{calc}$ 
($f_i^{fit}$) are the calculated (fitted) $S(E)$ results.
By inspection of the table, we conclude that
the values of $S^{n}(0)$ are strongly dependent
on the order of the polynomial function.  However,
an accurate description of the data can be obtained with
a desired degree of accuracy by increasing the 
number of polynomial terms.  With a cubic fit, for instance,
$\chi^2\sim 10^{-4}$ indicates that the calculated $S(E)$ values
are nicely reproduced.  This can be appreciated also in Fig.~\ref{fig:fit},
where the cubic fit is compared with the results for $S(E)$
obtained retaining all $(S+P)$-waves or only the
$^1S_0$ channel, using $\Lambda=500$ MeV with one particular
value of $c_D$ ($c_D=-0.20$).
The curve obtained using Eq.~(\ref{eq:taylor}) with the values for
$S(0)$, $S^{n=1}(0)$ and $S^{n=2}(0)$ of
Ref.~\cite{Che12} is also shown.
For energies up to 15 keV, the differences between
our $^1S_0$ results and those of Ref.~\cite{Che12}
are very small.  However, at energies of 25--30 keV
or higher, the quadratic fit of Ref.~\cite{Che12} starts to 
be significantly different from the calculated values,
as well as from the cubic fit.

Finally, using the results corresponding to the cubic fit in 
Table~\ref{tab:fit}, we have calculated that
the linear and quadratic contributions to $S(E)$ at the
solar Gamow peak are of the order of 7\% and 0.5\%, respectively,
while the cubic one is negligible. This is in agreement with
Refs.~\cite{Bac69,Che12}.  On the other hand,
for larger-mass stars, whose central temperature is of the
order of $5\times 10^7$ K and the Gamow peak is at $E\sim 15$ keV,
the linear, quadratic and cubic contributions become of the order
of 18\%, 3\% and 0.7\%, respectively.

The work of R.S. is supported by the U.S. Department of Energy,
Office of Nuclear Science, under contract DE-AC05-06OR23177.


\vspace{-0.1cm}
\begin{thebibliography}{100}
%
\bibitem{SFII}
E.G.\ Adelberger {\it et al.},
Rev.\ Mod.\ Phys.\ {\bf 83}, 195 (2011).
%
\bibitem{Bac69}
J.N.\ Bahcall and R.M.\ May, 
Astrphys.\ J.\ {\bf 155}, 511 (1969).
%
\bibitem{Sch98}
R.\ Schiavilla {\it et al.}, 
Phys.\ Rev.\ C {\bf 58}, 1263 (1998).
%
\bibitem{Che12}
J.-W.\ Chen, C.-P.\ Liu, and S.-H.\ Yu,
arXiv:1209.2552
%
\bibitem{Ent03}
D.R.\ Entem and R.\ Machleidt,
Phys.\ Rev.\ C {\bf 68}, 041001 (2003).
%
\bibitem{Mac11}
R.\ Machleidt and D.R.\ Entem,
Phys.\ Rep.\ {\bf 503}, 1 (2011).
%
\bibitem{Wir95}
R.B.\ Wiringa, V.G.J.\ Stoks, and R.\ Schiavilla,
Phys.\ Rev.\ C {\bf 51}, 38 (1995).
%
\bibitem{Par96ecc}
T.-S.\ Park, D.-P.\ Min, and M.\ Rho,
Nucl.\ Phys.\ A {\bf 596}, 515 (1996);
Y.-H.\ Song, R.\ Lazauskas, and T.-S.\ Park,
Phys.\ Rev.\ C {\bf 79}, 064002 (2009);
T.-S.\ Park {\it et al.},
Phys.\ Rev.\ C {\bf 67}, 055206 (2003).
%
\bibitem{Mar12}
L.E.\ Marcucci, A.\ Kievsky, S.\ Rosati, R.\ Schiavilla,
and M.\ Viviani,
Phys.\ Rev.\ Lett.\ {\bf 108}, 052502 (2012).
%
\bibitem{Marinprep}
L.E.\ Marcucci, R.\ Schiavilla, M.\ Viviani, in preparation.
%
\bibitem{She12}
G.\ Shen, L.E.\ Marcucci, J.\ Carlson, S.\ Gandolfi,
and R.\ Schiavilla,
Phys.\ Rev.\ C {\bf 86}, 035503 (2012).
%
\bibitem{Kur03}
A.\ Kurylov, M.J.\ Ramsey-Musolf, and P.\ Vogel,
Phys.\ Rev.\ C {\bf 67}, 035502 (2003).
%
\bibitem{Har90}
J.C.\ Hardy, I.S.\ Towner, V.T.\ Koslowsky, E.\ Hagberg, and H.\ Schmeing,
Nucl.\ Phys.\ A {\bf 509}, 429 (1990).
%
\bibitem{Mar00}
L.E.\ Marcucci, R.\ Schiavilla, M.\ Viviani, A.\ Kievsky, and S.\ Rosati,
Phys.\ Rev.\ Lett.\ {\bf 84}, 5959 (2000);
L.E.\ Marcucci {\it et al.}, 
Phys.\ Rev.\ C {\bf 63}, 015801 (2000).
%
\bibitem{Mar11}
L.E.\ Marcucci {\it et al.},
Phys.\ Rev.\ C {\bf 83}, 014002 (2011).
%
%
%
\end{thebibliography}
\end{document}